\documentclass[11pt]{article}
\usepackage[PostScript=dvips]{diagrams}
\usepackage{latexsym}
\usepackage{amssymb}
\usepackage{pstricks,pst-node,pst-tree}

\title{A Structural Approach to Reversible Computation}

\author{Samson Abramsky\\
Oxford University Computing Laboratory}
\date{}

\newcommand{\linimpl}{\multimap}
\newcommand{\llbang}{\, !}

\newcommand{\Nat}{\ensuremath{\mathbb{N}}}

\newlength{\sqpreordheight}
\newlength{\sqpreorddepth}
\newcommand{\sqpreord}%
{\mathbin{\raisebox{-1.02ex}[\sqpreordheight][\sqpreorddepth]%
{$\stackrel{\textstyle \sqsubset}{\sim}$}}}
\newcommand{\sqgtpreord}%
{\mathbin{\raisebox{-1.02ex}[\sqpreordheight][\sqpreorddepth]%
{$\stackrel{\textstyle \sqsupset}{\sim}$}}}

\newcommand{\eqdef}{\equiv}

\newcommand{\proof}{\textsc{Proof}}
\newcommand{\qed}{\Box}

\newtheorem{theorem}{Theorem}[section]

\newtheorem{definition}{Definition}[section]
\newtheorem{proposition}{Proposition}[section]

\newcommand{\TSig}{T_{\Sigma}}
\newcommand{\TSigX}{\TSig (X)}
\newcommand{\Unif}[3]{\mathcal{U}(#1 , #2 ){\downarrow} #3}
\newcommand{\NUnif}[2]{\mathcal{U}(#1 , #2 ){\uparrow}}
\newcommand{\labarrow}[1]{\stackrel{#1}{\longrightarrow}}

\newcommand{\opp}[1]{#1^{\mathsf{op}}}
\newcommand{\qinit}{q_{\iota}}
\newcommand{\qfin}{q_{f}}
\newcommand{\Aut}{\mathcal{A}}
\newcommand{\BAut}{\mathcal{B}}
\newcommand{\pinit}{p_{\iota}}
\newcommand{\pfin}{p_{f}}
\newcommand{\ap}{\cdot}
\newcommand{\LApp}[2]{\mathsf{LApp}(#1 , #2 )}
\newcommand{\Internal}[1]{#1^{\mathsf{int}}}
\newcommand{\Left}{l}
\newcommand{\Right}{r}
\newcommand{\Binop}{p}
\newcommand{\const}{\varepsilon}
\newcommand{\KK}{\mathbf{K}}
\newcommand{\SSS}{\mathbf{S}}
\newcommand{\Inj}{\mathcal{I}}
\newcommand{\PInv}{\mathcal{P}}
\newcommand{\CL}{\mathbf{CL}}
\newcommand{\BBB}{\mathbf{B}}
\newcommand{\CCC}{\mathbf{C}}
\newcommand{\DD}{\mathbf{D}}
\newcommand{\FF}{\mathbf{F}}
\newcommand{\WW}{\mathbf{W}}
\newcommand{\WWl}{\mathbf{W}}
\newcommand{\babs}[2]{\lambda^{\ast} #1 . \, #2}
\newcommand{\II}{\mathbf{I}}
\newcommand{\pfun}{\rightharpoonup}
\newcommand{\true}{\mathsf{true}}
\newcommand{\false}{\mathsf{false}}

\newcommand{\inp}{\mathsf{in}}
\newcommand{\out}{\mathsf{out}}
\newcommand{\labmap}[1]{\stackrel{#1}{\longmapsto}}

\newlength{\boxitwd} \newlength{\boxitht}% A couple of lengths
\newlength{\boxitdp} \newlength{\boxittotal}% A couple more. 
\newlength{\boxittitlewd} % And another
\newsavebox{\boxitcontents}        % and a box to save things in 
\newcommand{\boxit}[2][{}]{%
\savebox{\boxitcontents}{%                This box is the contents
\makebox[\textwidth]{%                    of the Pitts-style box
\begin{minipage}{0.95\textwidth}
{#2}
\end{minipage}
}}%
\settowidth{\boxittitlewd}{\hbox{#1}}%
\settowidth{\boxitwd}{\usebox{\boxitcontents}}%
\settoheight{\boxitht}{\usebox{\boxitcontents}}%
\settodepth{\boxitdp}{\usebox{\boxitcontents}}%
\setlength{\boxittotal}{\boxitht}%
\addtolength{\boxittotal}{\boxitdp}%
\begin{center}
\makebox[\textwidth][l]{% Box the whole thing up 
\raisebox{\boxitht}{%
\makebox[0pt][l]{%                   The top line and its title
\makebox[\textwidth][l]{\rule{0.1\textwidth}{0.1ex}%
\ifdim\boxittitlewd>0pt% Is there a title? 
  \raisebox{-0.5ex}{\textbf{{\ #1\ }}}\hrulefill}}}% Yes, put it in.
\else%
  \hrulefill% Just a line if there's no title
\fi%
\raisebox{-\boxitdp}{%   The bottom line
\makebox[0pt][l]{% 
\makebox[\textwidth][l]{\hrulefill}}}%
\makebox[0pt][l]{%       The left hand line and contents, flush left.
\makebox[\textwidth][l]{%
\raisebox{-\boxitdp}{\rule{0.1ex}{\boxittotal}}%
\usebox{\boxitcontents}%
}}%
\makebox[0pt][l]{%    The right hand line, flush right. 
\makebox[\textwidth][r]{%
\raisebox{-\boxitdp}{\rule{0.1ex}{\boxittotal}}%
}}%
}%
\end{center}
}

\begin{document}

\maketitle
\begin{abstract}
Reversibility is a key issue in the interface between computation and 
physics, and of growing importance as miniaturization progresses
towards its physical limits. Most foundational work  on reversible computing to date
has focussed on simulations of low-level machine models. By contrast,
we develop a more structural approach. We show how high-level
functional programs can be mapped \emph{compositionally}
(\textit{i.e.} in a syntax-directed fashion) into a simple kind of
automata which are immediately seen to be reversible. The size of the 
automaton is linear in the size of the functional term. In
mathematical terms, we are building a concrete \emph{model} of functional 
computation. This construction stems directly from ideas arising in
Geometry of Interaction and Linear Logic---but can be understood
without any knowledge of these topics. In fact, it serves as an
excellent introduction to them. At the same time, an interesting
logical delineation between reversible and irreversible forms of
computation emerges from our analysis.
\end{abstract}
\section{Introduction}
The importance of reversibility in computation, for both foundational
and, in the medium term, for practical reasons, is by now well
established. We  quote from the excellent
summary in the introduction to the recent paper by Buhrman, Tromp and
Vit\'anyi \cite{BTV}:
\begin{quotation}
\textbf{Reversible Computation:} R. Landauer \cite{Lan} has demonstrated
that it is only the ``logically irreversible'' operations in a
physical computer that necessarily dissipate energy by generating a
corresponding amount of entropy for every bit of information that gets 
irreversibly erased; the logically reversible operations can in
principle be performed dissipation-free. Currently, computations are
commonly irreversible, even though the physical devices that execute
them are fundamentally reversible. At the basic level, however, matter 
is governed by classical mechanics and quantum mechanics, which are
reversible. This contrast is only possible at the cost of efficiency
loss by generating thermal entropy into the environment. With
computational device technology rapidly approaching the elementary
particle level it has been argued many times that this effect gains in 
significance to the extent that efficient operation (or operation at
all) of future computers requires them to be reversible \ldots The
mismatch of computing organization and reality will express itself in
friction: computers will dissipate a lot of heat unless their mode of
operation becomes reversible, possibly quantum mechanical. 
\end{quotation}

\noindent The previous approaches of
which we are aware (e.g. \cite{Lec,Ben73,Ben82}) proceed by showing that some standard, low-level,
irreversible computational model such as Turing machines can be
simulated by a reversible version of the same model.
Our approach is  more
``structural''. We firstly define a simple model of computation which
is directly reversible in a very strong sense---every automaton $\Aut$ 
in our model has a ``dual'' automaton $\opp{\Aut}$, defined quite
trivially from $\Aut$, whose computations are exactly the
time-reversals of the computations of $\Aut$. We then establish a connection to models of
\emph{functional computation}. We will show that our model gives rise 
to a \emph{combinatory algebra} \cite{HS}, and derive universality as an 
easy consequence. This method of establishing universality has
potential significance for the important issue of how to
\emph{program} reversible computations.
To quote from \cite{BTV} again:
\begin{quotation}
Currently, almost no algorithms and other programs are designed
according to reversible principles \ldots To write reversible programs 
by hand is unnatural and difficult. The natural way is to compile
irreversible programs to reversible ones.
\end{quotation}
Our approach can be seen as providing a simple, \emph{compositional}
(i.e. ``syntax-directed'') compilation 
from high-level functional programs into a reversible model of
computation. This offers a novel perspective on reversible computing.

Our approach also has conceptual interest in that our constructions,
while quite concrete, are based directly on ideas stemming from Linear 
Logic and Geometry of Interaction \cite{Gi87,Gi89,Gi90,Gi95,GLR,MR91,DR93,DR, BP01}, and developed in previous
  work by the present author and a number of colleagues \cite{nfgoi,gfc,Abr96,Abr97,AbLe,AbLo,AHS}.
Our work here can be seen as a  concrete manifestation of these more
abstract and foundational developments. However, no knowledge of
Linear Logic or Geometry of Interaction is required to read the
present paper. In fact, it might serve as an
introduction to these topics, from a very concrete point of view. At the same time, an interesting
logical delineation between reversible and irreversible forms of
computation emerges from our analysis.

\subsection*{Related Work}
Geometry of Interaction (GoI) was initiated by Girard in a sequence of papers \cite{Gi89,Gi90,Gi95}, and extensively developed by Danos, Regnier, Malacaria, Baillot, Padicini and others, see e.g. \cite{MR91,DR93,DR,BP01}.
In particular, Danos and Regnier developed a computational view of GoI. In \cite{DR} they gave a compositional translation of the $\lambda$-calculus into a form of reversible abstract machine. We also note the thesis work of Mackie \cite{Mac}, done under the present author's supervision, which develops a GoI-based implementation paradigm for functional programming languages.

The present paper further develops the connections between GoI as a mathematical model of computation, and computational schemes with an emphasis on reversibility. As we see it, the main contributions are as follows:
\begin{itemize}
\item Firstly, the approach in the present paper seems particularly simple and direct. As already mentioned, we believe it will be accessible even without any prior knowledge of GoI or Linear Logic. The basic computational formalism is related very directly to standard ideas in term-rewriting, automata and combinatory logic.
By contrast, much of the literature on GoI can seem forbiddingly technical and esoteric to outsiders to the field. Thus we hope that this paper may help to open up some of the ideas in this field to a wider community.
\item There are also some interesting new perspectives on the standard ideas, e.g. the idea of biorthogonal term-rewriting system, and of linear combinatory logic (which was introduced by the present author in \cite{AHS}).
\item From the point of view of GoI itself, there are also some novelties. In particular, we develop the reversible computational structure in a \emph{syntax-free} fashion. We consider a general `space' of reversible automata, and define a linear combinatory algebra structure on this universe, rather than pinning all constructions to an induction on a preconceived syntax.
This allows the resulting structure to be revealed more clearly, and  the definitions and results to be stated more generally.

We also believe that our descriptions of the linear combinators as automata, and of application and replication as  constructions on automata, give a particularly clear and enlightening perspective on this approach to reversible functional computation.

\item The discussion in section~7 of the boundary between reversible and irreversible computation, and its relationship to pure vs. applied functional calculi, and the multiplicative-exponential vs. additive levels of Linear Logic, seems of conceptual interest, and is surely worth further exploration.

\item The results in section~8 on universality, and the consequent (and somewhat surprising) non-closure under linear application of finitely describable partial involutions, give rise to an interesting, and apparently challenging, open problem on the characterization of the realizable partial involutions.

\end{itemize}

\section{The Computational Model}
We formulate our computational model as a kind  of automaton with some simple
term-rewriting capabilities. We assume familiarity with the very
basic notions of term rewriting, such as may be gleaned from the
opening pages of any of the standard introductory accounts \cite{DJ,Klo,BN}.
In particular, we shall assume familiarity with the notions of
\emph{signature} $\Sigma = (\Sigma_{n} \mid n \in \omega )$, and of
the term algebras $\TSig$ and $\TSigX$, of ground terms, and terms in a 
set of variables $X$, respectively. We will work exclusively with
\emph{finite} signatures $\Sigma$. We also assume familiarity with the 
notion of \emph{most general unifier}; given terms $t, u \in \TSigX$, we write $\Unif{t}{u}{\sigma}$
if $\sigma : X \longrightarrow \TSigX$ is the most general unifying
substitution of $t$ and $u$, and $\NUnif{t}{u}$ if $t$ and
$u$ cannot be unified.

\noindent We define a \emph{pattern-matching automaton} to be a
structure
\[ \Aut \;\; = \;\; (Q, \qinit , \qfin , R) \]
where $Q$ is a finite set of states, $\qinit$ and $\qfin$ are
distinguished initial and final states, and $R \; \subseteq \; Q \times
\TSigX \times \TSigX \times Q$ is a finite set of \emph{transition
  rules}, written
\[ \begin{array}{rcl}
(q_1 , r_1 ) & \rightarrow & (s_1 , q_{1}' ) \\
& \vdots & \\
(q_{N}, r_N ) & \rightarrow & (s_N , q_{N}' )
\end{array} \]
where $q_{i}, q'_{i} \in Q$, $r_{i}, s_{i} \in \TSigX$, and the variables
occurring in $s_i$ are a subset of those occurring in $r_i$, $1
\leq i \leq N$. 
It is also convenient to assume that no variable appears in more than rule. 
We also stipulate that there are no incoming
transitions to the initial state, and no outgoing transitions from the 
final state: $\qinit \neq q'_{i}$ and $\qfin \neq q_i$, $1 \leq i \leq N$.

\noindent A \emph{configuration} of $\Aut$ is a pair $(q, t) \in Q
\times \TSig$ of a state and a ground term.
$\Aut$ induces a relation $\labarrow{\Aut}$ on configurations:
$(q, t) \labarrow{\Aut} (q',t')$ iff 
\[ \exists i  \,
(q_{i} = q \; \wedge \; q_{i}' = q' 
\; \wedge \;
\Unif{t}{r_{i}}{\sigma} \; \wedge \; t' = \sigma (s_{i}) ) . 
\]
Note that the ``pattern'' $r_i$ has to match the whole of the term $t$.
This is akin to the use of pattern-matching
in functional programming languages such as SML \cite{MTH} and Haskell
\cite{PJ98}, and is the reason for our choice of terminology.

Note that the cost of computing the transition relation $(q, t)
\labarrow{\Aut} (q',t')$ is \emph{independent} of the size of the ``input''
term $t$.\footnote{Under the assumption of \emph{left-linearity} (see
  below) which we shall shortly make, and on the standard assumption made in the
  algorithmics of unification \cite{BN,DJ} that the immediate sub-terms of
  a given term can be accessed in constant time.}
If we are working with a fixed pattern-matching automaton $\Aut$, this 
means that
the basic computation steps can be performed in constant time and
space, indicating that our computational model is at a reasonable
level of granularity.

A \emph{computation} over $\Aut$ starting with
an initial ground term $t_0 \in \TSig$ (the \emph{input}) is a sequence
\[ (\qinit , t_0 ) \labarrow{\Aut} (q_{1},t_{1}) \labarrow{\Aut} \cdots . \]
The computation is \emph{successful} if it terminates in a
configuration
$(\qfin ,t_{k})$, in which case $t_k$ is the
\emph{output}. Thus we can see a pattern-matching automaton as a
device for computing relations on ground terms.

\noindent We say that a pattern-matching automaton
\[ \Aut \;\; = \;\; (Q, \qinit , \qfin , R) \]
with
\[ R = \{ (q_{i}, r_{i}) \rightarrow (s_{i}, q_{i}') \mid 1 \leq i
\leq N \} \]
is \emph{orthogonal} if the
following conditions hold:
\begin{description}
\item[Non-ambiguity] For each $1 \leq i < j \leq N$, if $q_i = q_j$, then
  $\NUnif{r_{i}}{r_{j}}$.
\item[Left-linearity] For each $i$, $1 \leq i \leq N$, no
  variable occurs more than once in $r_i$.
\end{description}
Note that non-ambiguity is stated in a simpler form than the standard
version for term-rewriting systems \cite{BN,DJ,Klo}, taking advantage of the
fact that we are dealing with the simple case of pattern-matching.

Clearly the effect of non-ambiguity is that computation is \emph{deterministic}: given a
configuration $(q, t)$, at most one transition rule is applicable, so that the relation
$\labarrow{\Aut}$ is a partial function.

\noindent Given a pattern matching automaton $\Aut$ as above,
we define $\opp{\Aut}$ to be 
\[ (Q, \qfin , \qinit , \opp{R}) \]
where
\[ \opp{R} = \{ (q_{i}', s_{i}) \rightarrow (r_{i}, q_{i}) \mid 1 \leq i
\leq N \} \]
We define $\Aut$ to be \emph{biorthogonal} if both $\Aut$ and $\opp{\Aut}$ are
orthogonal pattern-matching automata.
Note that if $\Aut$ is a biorthogonal automaton, so is $\opp{\Aut}$,
and $\Aut^{\mathsf{op} \, \mathsf{op}} = \Aut$.

It should be clear that computation in biorthogonal automata is
reversible in a deterministic, step-by-step fashion. Thus if we have
the computation
\[ (\qinit ,t_{0}) \labarrow{\Aut} \cdots \labarrow{\Aut} (\qfin ,t_{n}) \]
in the biorthogonal automaton $\Aut$, then we have the computation
\[  (\qfin ,t_{n}) \labarrow{\opp{\Aut}} \cdots \labarrow{\opp{\Aut}} (\qinit ,t_{0}) \]
in the biorthogonal automaton $\opp{\Aut}$. Note also that biorthogonal
automata are \emph{linear} in the sense that, for each rule $(q, r)
\rightarrow (s, q')$, the same variables occur in $r$ and in $s$, and
moreover each variable which occurs does so exactly once in $r$
and exactly once in $s$. Thus there is no ``duplicating'' or
``discarding'' of sub-terms matched to variables in applying a rule,
whether in $\Aut$ or in $\opp{\Aut}$.

Orthogonality is a very standard and important condition in
term-rewriting systems. However, biorthogonality is a much stronger
constraint, and very few of the term-rewriting systems usually
considered satisfy this condition.
(In fact, the only familiar examples of biorthogonal rewriting systems 
seem to be associative/commutative rewriting and similar, and these are
usually considered as notions for ``rewriting modulo'' rather than as
computational rewriting systems in their own right).

Our model of computation will be the class of biorthogonal
pattern-matching automata; from now on, these will be the only automata
we shall consider, and we will refer to them simply as ``automata''.
The reader will surely agree that this computational model is quite
simple, and seen to be reversible in a very direct and immediate
fashion. We will now turn to the task of establishing its
universality.

\paragraph{Remark}
It would have been possible to represent our computational model more
or less entirely in terms of standard notions of term rewriting
systems. We briefly sketch how this might be done. Given an automaton
\[  \Aut \;\; = \;\; (Q, \qinit , \qfin , R) \]
we expand the (one-sorted) signature $\Sigma$ to a signature over three
sorts: $V$ (for values), $S$ (for states) and $C$ (for
configurations). The operation symbols in $\Sigma$ have all their
arguments and results of sort $V$; for each state $q \in Q$, there is
a corresponding constant of sort $S$; and there is a binary operation
\[ \langle \cdot , \cdot \rangle : S \times V \longrightarrow C . \]
Now the transition rules $R$ turn into a rewriting system in the
standard sense; and orthogonality has its standard meaning. We would
still need to focus on initial terms of the form $\langle \qinit , t
\rangle$ and normal forms of the form $\langle \qfin , t
\rangle$, $t$ ground.

Our main reason for using the automaton formulation is that it does
expose some salient structure, which will be helpful in defining and
understanding the significance of the constructions to follow.

\section{Background on Combinatory Logic}
In this section, we briefly review some basic material. For further details, see
\cite{HS}.

We recall that combinatory logic is the algebraic theory $\CL$ given
by the signature with one binary operation (application) written
as an infix $\_\cdot\_$, and two constants $\SSS$ and $\KK$, subject to 
the equations
\[ \begin{array}{lcl}
\KK \cdot x \cdot y & = & x \\
\SSS \cdot x \cdot y \cdot z & = & x \cdot z \cdot (y \cdot z) 
\end{array} \]
(application associates to the left, so $x \cdot y \cdot z = (x \cdot
y) \cdot z$). Note that we can define $\II \equiv \SSS \cdot \KK \cdot
\KK$, and verify that $\II \cdot x = x$.

The key fact about the combinators is that they are \emph{functionally 
  complete}, i.e. they can simulate the effect of
$\lambda$-abstraction. Specifically, we can define bracket abstraction 
on terms in $T_{\CL}(X)$:
\[ \begin{array}{lcl}
\babs{x}{M} & = & \KK \cdot M \quad (x \not\in \mathsf{FV}(M)) \\
\babs{x}{x} & = & \II  \\
\babs{x}{M \cdot N} & = & \SSS \cdot (\babs{x}{M}) \cdot (\babs{x}{N}) 
\end{array} \]
Moreover (Theorem 2.15 in \cite{HS}):
\[ \CL \vdash (\babs{x}{M}) \cdot N = M[N/x]. \]

\noindent The $\BBB$ combinator can be defined by bracket abstraction from its
defining equation:
\[ \BBB \cdot x \cdot y \cdot z = x \cdot (y \cdot z) .\]
The combinatory \emph{Church numerals} are then defined by
\[ \bar{n} \equiv (\SSS \cdot \BBB )^{n} \cdot (\KK \cdot \II ) \]
where we define
\[ a^n \cdot b = a\cdot (a \cdots (a \cdot b)\cdots ) . \]
A partial function $\phi : \Nat \pfun \Nat$ is \emph{numeralwise
  represented} by a combinatory term $M \in T_{\CL}$ if for all $n \in \Nat$, if
$\phi (n)$ is defined and equal to $m$, then
\[ \CL \vdash M \cdot \bar{n} = \bar{m} \]
and if $\phi (n)$ is undefined, then $M\cdot \bar{n}$ has no normal
form.

\noindent The basic result on computational universality of $\CL$ is then the following (Theorem 4.18 in \cite{HS}):
\begin{theorem}
\label{clcomp}
The  partial functions numeralwise representable in $\CL$ are exactly
the partial recursive functions.
\end{theorem}

\section{Linear Combinatory Logic}

We shall now present another system of combinatory logic: \emph{Linear 
  Combinatory Logic} \cite{Abr97,AHS,AbLe}. This can be seen as a finer-grained system into
which standard combinatory logic, as presented in the previous
section, can be interpreted. By exposing some finer structure, Linear
Combinatory Logic offers a more accessible and insightful path towards 
our goal of mapping functional computation into our simple model of
reversible computation.

Linear Combinatory Logic can be seen as the combinatory analogue of
Linear Logic \cite{Gi87}; the interpretation of standard Combinatory Logic into
Linear Combinatory Logic corresponds to the interpretation of
Intuitionistic Logic into Linear Logic. Note, however, that the
combinatory systems we are considering are type-free and
``logic-free'' (\textit{i.e.} purely equational).

\begin{definition}\label{LCA}
A {\em Linear Combinatory Algebra} $(A,\ap, \llbang)$ consists of the following 
data:
\begin{itemize}
\item An applicative structure $(A,\ap)$
\item A unary operator $\llbang: A\rightarrow A$
\item Distinguished elements $\BBB$, $\CCC$, $\II$, $\KK$, 
  $\DD$, $\delta$, $\FF$, $\WW$ of $A$
\end{itemize}
satisfying the following identities (we associate $\ap$ to the left and write
$ x  \, \cdot \llbang y$ for $x
\ap (\llbang (y))$, etc.) for all variables $x,y,z$ ranging over $A$.
\[ \begin{array}{clclr}
1. &  \BBB \cdot x \cdot y \cdot z & = &  x \cdot (y \cdot z) &
\mbox{Composition/Cut} \\
2. & \CCC \cdot x \cdot y \cdot z & = &  (x \cdot z) \cdot y & \mbox{Exchange} \\
3. & \II \cdot x & = &  x & \mbox{Identity} \\
4. & \KK \cdot x \, \cdot \llbang y & = &  x & \mbox{Weakening} \\
5. & \DD \, \cdot \llbang x & = &  x & \mbox{Dereliction} \\
6. & \delta \, \cdot \llbang x & = &  \llbang \llbang x &
  \mbox{Comultiplication} \\
7. & \FF \, \cdot \llbang x \, \cdot \llbang y & = &  \llbang (x \cdot
  y) & \mbox{Monoidal Functoriality} \\
8. & \WW  \cdot x \, \cdot \llbang y & = &  x \, \cdot \llbang y \, \cdot
  \llbang y & \mbox{Contraction} \\
\end{array} \]
\end{definition}
 
\noindent
The notion of LCA corresponds to a Hilbert style
axiomatization of the  $\{!, \linimpl\}$ fragment of
linear logic \cite{Abr97,Avr88,Tro92}. The \emph{principal types} of
the combinators correspond to the axiom schemes which they name. They
can be computed by a Hindley-Milner style algorithm \cite{Hin} from the
above equations:
\[ \begin{array}{llcl}
1. & \BBB & : & (\beta \linimpl \gamma ) \linimpl (\alpha \linimpl \beta ) \linimpl
  \alpha \linimpl \gamma \\
2. & \CCC & : & (\alpha \linimpl \beta \linimpl \gamma ) \linimpl (\beta \linimpl \alpha 
  \linimpl \gamma ) \\
3. & \II & : & \alpha \linimpl \alpha \\
4. & \KK & : & \alpha \linimpl \llbang \beta \linimpl \alpha \\
5. & \DD & : & \llbang \alpha \linimpl \alpha \\
6. & \delta & : & \llbang \alpha \linimpl \llbang \llbang \alpha \\
7. & \FF & : & \llbang (\alpha \linimpl \beta ) \linimpl \llbang \alpha
  \linimpl  \llbang 
\beta \\
8. & \WW & : & (\llbang \alpha \linimpl \llbang \alpha \linimpl \beta ) \linimpl \llbang \alpha
  \linimpl \beta \\
\end{array} \]
Here $\linimpl$ is a \emph{linear function type} (linearity means that 
the argument is used exactly once), and $\llbang \alpha$ allows
arbitrary copying of an object of type $\alpha$.

A {\em Standard Combinatory Algebra}  consists of a pair
$(A,\ap_s)$ where $A$ is a nonempty set and $\ap_s$ is a binary
operation on $A$, together with  distinguished
elements $\BBB_s, \CCC_s, \II_s, \KK_s,$ and $\WW_s$  of $A$,  satisfying
the following identities for all $x,y,z$ ranging over $A$:
\[ \begin{array}{llcl}
1. & \BBB_s\ap_s x\ap_s y \ap_s z & = &  x\ap_s (y\ap_s z) \\
2. & \CCC_s\ap_s x\ap_s y\ap_s z & = &  (x\ap_s z)\ap_s y \\
3. & \II_s \ap_s x & = &  x \\
4. & \KK_s\ap_s x\ap_s  y & = &  x  \\
5. & \WW_s \ap_s x \ap_s y & = &  x\ap_s y\ap_s y \\
\end{array} \]

\noindent Note that this is equivalent to the more familiar definition of
$\mathbf{SK}$-combinatory algebra as given in the previous section.
In particular, $\SSS_s$ can be defined from $\BBB_s$,
$\CCC_s$, $\II_s$   
and $\WW_s$ \cite{Bar84,Hin}. 
Let $(A,\ap,\llbang)$ be a linear combinatory algebra. We define a binary
operation $\ap_s$ on $A$ as follows: for $a, b \in A$,
$a \ap_s b \eqdef a \, \ap \llbang b$. 
We define $\DD' \eqdef \mathbf{C \cdot (B \cdot B \cdot I) \cdot (B \cdot D \cdot I)}$. Note that
\[ \DD'  \cdot x  \, \cdot \llbang y = x \cdot y. \]
Now consider
the following elements of $A$.
 
\[ \begin{array}{llcl}
1. & \BBB_s & \eqdef & \mathbf{C\ap (B\ap (B\ap B\ap B)\ap (D'\ap I))\ap (C\ap ((B\ap B)\ap F
)\ap \delta)} \\
2. & \CCC_s & \eqdef & \DD'\ap \CCC \\
3. & \II_s & \eqdef & \DD'\ap \II \\
4. & \KK_s & \eqdef & \DD'\ap \KK \\
5. & \WW_s & \eqdef & \DD'\ap \WW \\
\end{array} \]
 
\begin{theorem}\label{main2}
Let $(A,\ap,\llbang)$ be a linear combinatory algebra. Then $(A, \ap_s)$
with $\ap_s$ and the elements $\BBB_s, \CCC_s, \II_s, \KK_s, \WW_s$ as defined above is a 
standard combinatory algebra.
\end{theorem}

Finally, we mention a special case which will arise in our reversible
model.
An \emph{Affine Combinatory Algebra} is a Linear Combinatory Algebra
such that the $\KK$ combinator satisfies the stronger equation
\[ \KK \cdot x \cdot y = x. \]
Note that in this case we can \emph{define} the identity combinator: $\II
\eqdef \CCC \cdot \KK \cdot \KK$.

\section{The Affine Combinatory Algebras  $\mathcal{I}$ and $\mathcal{P}$}

We fix the following signature $\Sigma$ for the remainder of this
paper.
\[ \begin{array}{lcl}
\Sigma_0 & = & \{ \const \} \\
\Sigma_1 & = & \{ \Left , \Right \} \\
\Sigma_2 & = & \{ \Binop \} \\
\Sigma_n & = & \varnothing , \quad n > 2 .
\end{array} \]
We shall discuss minimal requirements on the signature in Section~6.4.

\noindent We write $\Inj$ for the set of all partial injective 
functions on $\TSig$.

\subsection{Operations on $\Inj$}

\subsubsection{Replication}
\[ \llbang f = \{ (p(t,u),p(t,v)) \mid t \in \TSig \; \wedge \; (u,v) \in f \} \]
\subsubsection{Linear Application}
\[ \LApp{f}{g} = f_{rr} \cup f_{rl} ; g ; (f_{ll} ; g)^{\ast} ; f_{lr} 
\]
where
\[ f_{ij} = \{ (u,v) \mid (i(u),j(v)) \in f \} \quad (i, j \in \{ l, r 
\} ) \]
and we use the operations of relational algebra (union, composition,
and reflexive, transitive closure).

The idea is that terms of the form $r(t)$ correspond to interactions
between the functional process represented by $f$ and its
environment, while terms of the form $l(t)$ correspond to interactions 
with its argument, namely the functional process represented by
$g$.
This is \emph{linear} application because the function interacts with
one copy of its argument, whose state changes as the function
interacts with it; ``fresh'' copies of the argument are not
necessarily available
as the computation proceeds. The purpose of the replication operation
described previously is precisely to make the argument copyable, using 
the first argument of the constructor $p$ to ``tag'' different copies.

The ``flow of control'' in linear application is indicated by the
following diagram:

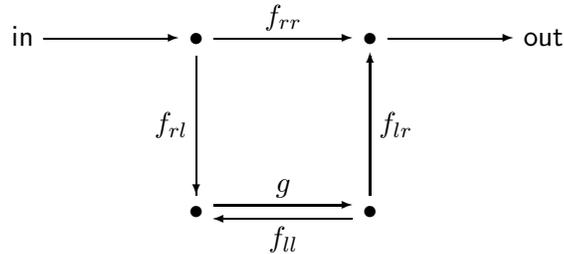
\[ \begin{diagram}
\mathsf{in} & \rTo & \bullet & \rTo^{f_{rr}} & \bullet  & \rTo &
\mathsf{out} \\
& & \dTo^{f_{rl}} & & \uTo_{f_{lr}} & & \\
& & \bullet & \pile{\rTo^{g} \\ \lTo_{f_{ll}}} & \bullet & & \\
\end{diagram} \]
Thus the function $f$ will either respond immediately to a request
from the environment without consulting its argument ($f_{rr}$), or it
will send a ``message'' to its argument ($f_{rl}$), which initiates a
dialogue between $f$ and $g$ ($f_{ll}$ and $g$), which ends with $f$
despatching a response to the environment ($f_{lr}$). This protocol is
mediated by the top-level constructors $l$ and $r$, which are used
(and consumed) by the operation of Linear Application.
 
\subsection{Partial Involutions}
Note that $f \in \Inj \; \Rightarrow \; \opp{f} \in \Inj$, where
$\opp{f}$ is the relational converse of $f$.
We say that $f \in \Inj$ is a \emph{partial involution} if $\opp{f} =
f$.
We write $\PInv$ for the set of partial involutions.

\begin{proposition}
Partial involutions are closed under replication and linear application.
\end{proposition}
\noindent \proof \ \  It is immediate that partial involutions are
closed under replication. Suppose that $f$ and $g$ are partial
involutions, and that $\LApp{f}{g}(u) = v$. We must show that
$\LApp{f}{g}(v) = u$. There are two cases.

\noindent Case 1: $f(r(u)) = r(v)$, in which case $f(r(v)) = r(u)$,
and $\LApp{f}{g}(v) = u$ as required.

\noindent Case 2: for some $w_1$, \ldots , $w_k$, $k \geq 0$,
\[ \begin{array}{l}
f(r(u )) = l(w_1 ), g(w_1 ) = w_2 , f(l(w_{2})) = l(w_{3}),
g(w_{3}) = w_4 , \ldots , f(l(w_{k} )) = l(w_{k+1}) , \\
g(w_{k+1}) = w_{k+2} , f(l(w_{k+2}) = r(v) . 
\end{array} \]
Since $f$ and $g$ are involutions, this implies
\[ \begin{array}{l}
f(r(v )) = l(w_{k+2} ), g(w_{k+2} ) = w_{k+1} , f(l(w_{k+1})) = l(w_{k}),
 \ldots , g(w_{4}) = w_3 , f(l(w_{3} )) = l(w_{2}) , \\
g(w_{2}) = w_{1} , f(l(w_{1}) = r(u) , 
\end{array} \]
and hence  $\LApp{f}{g}(v) = u$ as required.  $\;\;\; \qed$

\subsection{Realizing the linear combinators by partial involutions}

A partial injective map $f \in \Inj$ is \emph{finitely describable} if there is a family
\[ \{ (t_i , u_i ) \mid 1 \leq i \leq k \} \]
where $t_i , u_i \in \TSigX$,
such that the graph of $f$ is the symmetric closure of
\[ \{ (\sigma (t_i ), \sigma (u_i )) \mid \sigma : X \longrightarrow \TSig , \, 1 \leq i \leq k \} . \]
Here $\sigma : X \longrightarrow \TSig$ ranges over \emph{ground substitutions}.

We write $t \leftrightarrow u$ when $(t, u)$ is in the finite description of a partial involution, and refer to such expressions as \emph{rules}.

\subsubsection{The identity combinator $\II$}
As a first, very simple case, consider the identity combinator $\II$, 
with the defining equation
\[ \II \cdot a = a . \]
We can picture the $\II$ combinator, which should evidently be applied 
to one argument to achieve its intended effect, thus:
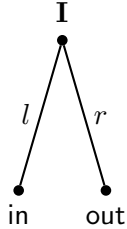
\psset{labelsep=2pt,radius=2pt,tnpos=a,tnsep=1pt}
\begin{center}
\pstree{\TC*~{$\II$}}{
\TC*~[tnpos=b]{$\inp$}\tlput{$\Left$}
\TC*~[tnpos=b]{$\out$}\trput{$\Right$}
}
\end{center}
Here the tree represents the way the applicative structure is encoded
into the constructors $\Left$, $\Right$, as reflected in the
definition of $\mathsf{LApp}$. Thus when $\II$ is applied to an argument $a$,
the $\Left$-branch will be connected to $a$, while the $\Right$-branch 
will be connected to the output. The equation $\II \cdot a = a$ means 
that we should have \emph{the same information} at the leaves $a$ and $\out$
of the tree. This can be achieved by the rule
\[ \II : \quad \Left (x) \leftrightarrow \Right (x)  \]
and this yields the definition of the automaton for $\II$.

Now we can show that for any automaton $\Aut$ representing an argument
$a$ we indeed have
\[ f_{\LApp{\Aut_{\II}}{\Aut}} = f_{\Aut} = a . \]
Indeed, for any input $t$
\[ \frac{\Right (t) \labmap{\II} \Left (t) \quad t \labmap{f} u \quad
\Right (u) \labmap{\II} \Left (u)}{t \labmap{\LApp{\Aut_{\II}}{\Aut}} u} 
\]

\subsubsection{The constant combinator $\KK$}
Now we consider the combinator $\KK$, with the defining equation, with defining equation
\[ \KK \cdot a \cdot b = a . \]
We have the tree diagram
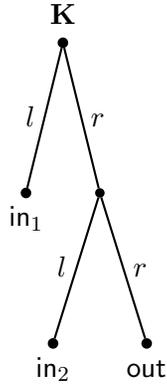
\begin{center}
\pstree{\TC*~{$\KK$}}{
\TC*~[tnpos=b]{$\inp_1$}\tlput{$\Left$}
\pstree{\Tdot\trput{$\Right$}}
 {\TC*~[tnpos=b]{$\inp_2$}\tlput{$\Left$}
  \TC*~[tnpos=b]{$\out$}\trput{$\Right$}
 }
}
\end{center}
The defining equation means that we need to make the information at $\out$ equalt to that at $\in_1$. This can be accomplished by the rule
\[ \KK : \Left (x) \leftrightarrow \Right (\Right (x)) . \]
Note that the second argument ($\in_2$) does not get accessed by this rule.

\subsubsection{The bracketing combinator $\BBB$}
We now turn to a more complex example, the `bracketing' combinator $\BBB$, with the defining equation
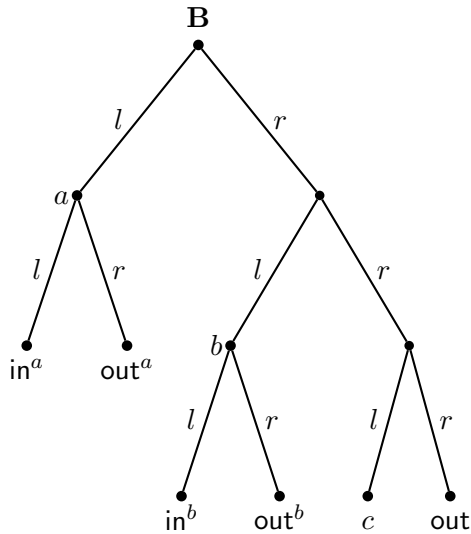
\[ \BBB \cdot a \cdot b \cdot c = a \cdot (b \cdot c) . \]
\begin{center}
\pstree{\TC*~{$\BBB$}}{
\pstree{\TC*~[tnpos=l]{$a$}\tlput{$\Left$}}
 {\TC*~[tnpos=b]{$\inp^a$}\tlput{$\Left$}
  \TC*~[tnpos=b]{$\out^a$}\trput{$\Right$}
 }
\pstree{\Tdot\trput{$\Right$}}
 {\pstree{\TC*~[tnpos=l]{$b$}\tlput{$\Left$}}
   {\TC*~[tnpos=b]{$\inp^b$}\tlput{$\Left$}
    \TC*~[tnpos=b]{$\out^b$}\trput{$\Right$}
   }
  \pstree{\Tdot\trput{$\Right$}}
   {\TC*~[tnpos=b]{$c$}\tlput{$\Left$}
    \TC*~[tnpos=b]{$\out$}\trput{$\Right$}
   }
 }
}
\end{center}
Here, the arguments $a$ and $b$ themselves have some applicative structure used in the defining equation: $a$ is applied to the rsult of applying $b$ to $c$. This means that the automaton realizing $\BBB$ must access the argument and result positions of $a$ and $b$, as shown in the tree diagram.

This requires the output $\out$ of $\BBB$ to be connected to the output $\out^a$ of $a$.
This translates into the following rule:
\[ \Right (\Right (\Right (x))) \leftrightarrow \Left (\Right (x)) . \]
Similarly, the output $\out^b$ of $b$ must be connected to $\inp^a$, leading to the rule:
\[ \Left (\Left (x)) \leftrightarrow \Right (\Left (\Right (x))) . \]
Finally, $c$ must be connected to $\inp^b$, leading to the rule:
\[ \Right (\Left (\Left (x)) \leftrightarrow \Right (\Right (\Left (x))) . \]

\subsubsection{The commutation combinator $\CCC$}
The $\CCC$ combinator can be analyzed in a similar fashion. The defining equation is
\[ \CCC \cdot a \cdot b \cdot c = a \cdot c \cdot b . \]
We have the tree diagram
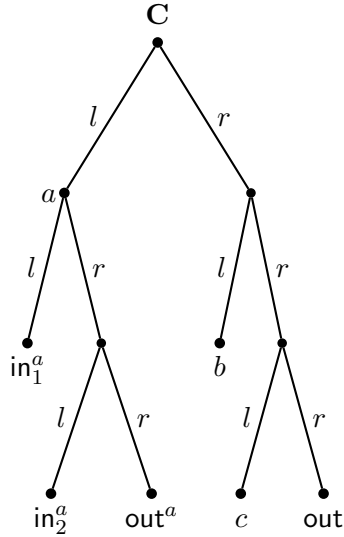
\begin{center}
\pstree{\TC*~{$\CCC$}}{
\pstree{\TC*~[tnpos=l]{$a$}\tlput{$\Left$}}
 {\TC*~[tnpos=b]{$\inp_1^a$}\tlput{$\Left$}
  \pstree{\Tdot\trput{$\Right$}}
   {\TC*~[tnpos=b]{$\inp_2^a$}\tlput{$\Left$}
    \TC*~[tnpos=b]{$\out^a$}\trput{$\Right$}
   }
 }
\pstree{\Tdot\trput{$\Right$}}
 {\TC*~[tnpos=b]{$b$}\tlput{$\Left$}
  \pstree{\Tdot\trput{$\Right$}}
   {\TC*~[tnpos=b]{$c$}\tlput{$\Left$}
    \TC*~[tnpos=b]{$\out$}\trput{$\Right$}
   }
 }
}
\end{center}
We need to connect $b$ to $\inp^a_2$, $c$ to $\inp^a_1$, (this inversion of the left-to-right ordering corresponds to the commutative character of this combinator), and $\out$ to $\out^a$. We obtain the following set of rules:
\[ R_{\CCC}:
\begin{array}{lcl}
l(l(x)) & \leftrightarrow & r(r(l(x))) \\
l(r(l(x))) & \leftrightarrow & r(l(x))) \\
l(r(r(x))) & \leftrightarrow & r(r(r(x)))
\end{array} \]

Note at this point that linear combinatory completeness already yields something rather striking in these terms; that all patterns of accessing arguments and results, with arbitrarily nested applicative stucture, can be generated by just the above combinators under linear application.

Note that at the multiplicative level, we only need unary operators in the term algebra. To deal with the exponential $\llbang$, a binary constructor is needed.

\subsubsection{The dereliction combinator $\DD$}
We start with the dereliction combinator $\DD$, with defining equation
\[ \DD \cdot \llbang a = a . \]
Notice that the combinator expects an argument of a certain form, namely $\llbang a$ (and the equational rule will only ``fire'' if it has that form).

We have the tree
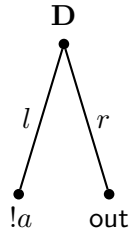
\begin{center}
\pstree{\TC*~{$\DD$}}{
\TC*~[tnpos=b]{$\llbang a$}\tlput{$\Left$}
\TC*~[tnpos=b]{$\out$}\trput{$\Right$}
}
\end{center}
We need to connect the output to \emph{one copy} of the input. We use the constant $\epsilon$ to pick out this copy, and obtain the rule:
\[ \Left (\Binop (\epsilon , x)) \leftrightarrow \Right (x) . \]

\subsubsection{The comultiplication combinator $\delta$}
For the comultiplication operator, we have the equation
\[ \delta \cdot \llbang a = \llbang \llbang a \]
and the tree
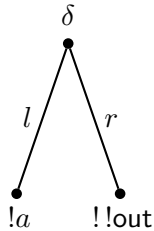
\begin{center}
\pstree{\TC*~{$\delta$}}{
\TC*~[tnpos=b]{$\llbang a$}\tlput{$\Left$}
\TC*~[tnpos=b]{$\llbang \llbang \out$}\trput{$\Right$}
}
\end{center}
Note that a typical pattern at the output will have the form
\[ \Right (\Binop (x, \Binop (y, z))) \]
while a typical pattern at the input has the form
\[ \Left (\Binop (x' , y')) . \]
The combinator cannot control the shape of the sub-term at $y'$, so we cannot simply unify the two patterns. However, because of the nature of the replication operator, we can impose whatever structure we like on the `copy tag' $x'$, in the knowledge that this will not be changed by the argument $\llbang a$ which the combinator will be applied to
Hence we can  match these two patterns up, using the fact that the term algebra $\TSig$ allows arbitrary nesting of constructors, so that we can write a pattern for the input as
\[ \Left (\Binop (\Binop (x, y), z)) . \]
Thus we obtain the rule
\[ \Left (\Binop (\Binop (x, y), z)) \leftrightarrow \Right (\Binop (x, \Binop (y, z))) . \]
Note that this rule embodies an ``associativity isomorphism for pairing'', although of course in the free term algebra $\TSig$ the constructor $p$ is certainly not associative.

\subsubsection{The functional distribution combinator $\FF$}
The combinator $\FF$ with equation
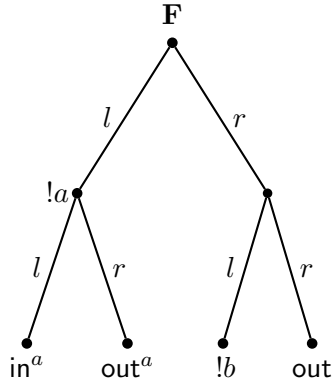
\[ \FF \cdot \llbang a \cdot \llbang b = \llbang (a \cdot b) . \]
\begin{center}
\pstree{\TC*~{$\FF$}}{
\pstree{\TC*~[tnpos=l]{$\llbang a$}\tlput{$\Left$}}
 {\TC*~[tnpos=b]{$\inp^a$}\tlput{$\Left$}
  \TC*~[tnpos=b]{$\out^a$}\trput{$\Right$}
 }
\pstree{\Tdot\trput{$\Right$}}
 {\TC*~[tnpos=b]{$\llbang b$}\tlput{$\Left$}
  \TC*~[tnpos=b]{$\out$}\trput{$\Right$}
 }
}
\end{center}
$\FF$ expresses `closed functoriality' of $\llbang$ with respect to the linear hom $\linimpl$.
Concretely, we must move the application of $a$ to $b$ inside the $\llbang$, which is achieved by commuting the constructors $\Left$, $\Right$ and $\Binop$. Thus we connect $\out^a$ to $\out$:
\[ \Left (\Binop (x, \Right (y))) \leftrightarrow \Right (\Right (\Binop (x, y))) \]
and $\inp^a$ to $\llbang b$:
\[ \Left (\Binop (x, \Left (y))) \leftrightarrow \Right (\Left (\Binop (x, y))) . \]

\subsubsection{The duplication combinator $\WWl$}
Finally, we consider the duplication combinator $\WWl$:
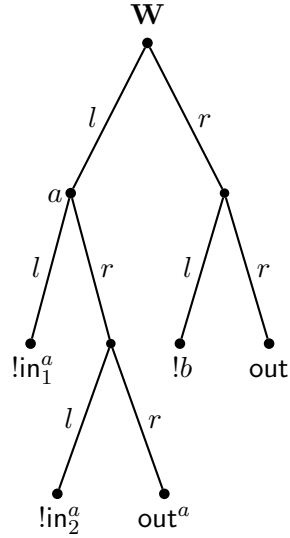
\[ \WWl \cdot a \cdot \llbang b = a \cdot \llbang b \cdot \llbang b . \]
\begin{center}
\pstree{\TC*~{$\WWl$}}{
\pstree{\TC*~[tnpos=l]{$a$}\tlput{$\Left$}}
 {\TC*~[tnpos=b]{$\llbang \inp_1^a$}\tlput{$\Left$}
  \pstree{\Tdot\trput{$\Right$}}
   {\TC*~[tnpos=b]{$\llbang \inp_2^a$}\tlput{$\Left$}
    \TC*~[tnpos=b]{$\out^a$}\trput{$\Right$}
   }
 }
\pstree{\Tdot\trput{$\Right$}}
 {\TC*~[tnpos=b]{$\llbang b$}\tlput{$\Left$}
  \TC*~[tnpos=b]{$\out$}\trput{$\Right$}
 }
}
\end{center}
We must connect $\out$ and $\out^a$:
\[ \Right (\Right (x)) \leftrightarrow \Left (\Right (\Right (x))) . \]
We also need to connect $\llbang b$ \emph{both} to $\inp^a_1$ \emph{and} to $\inp^a_2$. We do this by using the copy-tag field of $\llbang b$ to split its address space into two, using the constructors $\Left$ and $\Right$. This tag tells us whether a given copy of $\llbang b$ should be connected to the first ($\Left$) or second ($\Right$) input of $a$.
Thus we obtain the rules:
\[ \begin{array}{ccc}
\Left (\Left (\Binop (x, y))) & \leftrightarrow & \Right (\Left (\Binop (\Left (x), y))) \\
\Left (\Right (\Left (\Binop (x, y)))) & \leftrightarrow & \Right (\Left (\Binop (\Right (x), y)))
\end{array}
\]
Once again, combinatory completeness tells us that from this limited stock of combinators, \emph{all} definable patterns of application can be expressed; moreover, we have a universal model of computation.

\subsection{The affine combinatory algebras $\Inj$ and $\PInv$}
\begin{theorem}
\label{injth}
$(\Inj , \cdot , {\llbang}, f_{\BBB}, f_{\CCC}, f_{\KK}, f_{\DD},
f_{\delta}, f_{\FF}, f_{\WWl})$ is an affine combinatory
algebra, with subalgebra $\PInv$.
\end{theorem}
This theorem is a variation on the results established in
\cite{Abr96,Abr97,AbLo,AbLe,AHS}; see in particular
\cite[Propositions 4.2, 5.2]{AHS}, and the combinatory algebra of partial involutions studied
in \cite{AbLe}. The ideas on which this construction is based stem from
Linear Logic \cite{Gi87,GLR} and Geometry of Interaction \cite{Gi89,Gi90}, in the form
developed by the present author and a number of colleagues
\cite{nfgoi,gfc,Abr96,Abr97,AbLo,AbLe,AHS}.

Once again, combinatory completeness tells us that from this limited
stock of combinators, \emph{all} definable patterns of application can
be expressed; moreover, we have a universal model of computation.

\section{Automatic Combinators}

As we have already seen, a pattern-matching automaton $\Aut$ can be
seen as a device for computing a relation on ground terms. The
relation $R_{\Aut} \, \subseteq \, \TSig \times \TSig$ is the set of
all pairs $(t, t')$ such that there is a computation
\[ (\qinit , t) \labarrow{\Aut}^{\ast} (\qfin , t') . \]
In the case of a biorthogonal automaton $\Aut$, the relation
$R_{\Aut}$ is in fact a \emph{partial injective function}, which we
write $f_{\Aut}$. Note that $f_{\opp{\Aut}} = \opp{f_{\Aut}}$, the
converse of $f_{\Aut}$, which is also a partial injective function.  
In the previous section, we defined a linear combinatory algebra $\PInv$ based on the set of partial involutions on $\TSig$. 
We now want to define a subalgebra of $\PInv$ consisting of those partial
involutions ``realized'' or ``implemented'' by a biorthogonal
automaton.
We refer to such combinators as ``Automatic'', by analogy with
Automatic groups \cite{ECHL}, structures \cite{KN95} and sequences \cite{AS03}.

\subsection{Operations on Automata}
\subsubsection{Replication}
Given an automaton $\Aut = (Q, \qinit , \qfin , R)$, let $x$ be a
variable not appearing in any rule in $R$. We define
\[ \llbang \Aut = (Q, \qinit , \qfin , \llbang R) \]
where $\llbang R$ is defined as
\[  \{ (q, \Binop (x, r)) \rightarrow (\Binop (x,s),q' )
\mid (q, r) \rightarrow (s, q') \in R \} . \]
Note that the condition on $x$ is necessary to ensure the linearity of 
$\llbang R$. The biorthogonality of $\llbang \Aut$ is easily verified.

\subsubsection{Linear Application}
See Figure~1.
Here $Q \uplus P$ is the disjoint union of $Q$ and $P$ (we simply
assume that $Q$ and $P$ have been relabelled if necessary to be
disjoint).

\begin{figure*}
\[ 
\Aut \; = \; (Q, \qinit , \qfin , R) \qquad \qquad
\BAut \; = \; (P, \pinit , \pfin , S)
\]
\[ \LApp{\Aut}{\BAut} =
(Q \uplus P, \qinit , \qfin , T)  \]
\[ T = \bigcup_{j, k \in \{\Left , \Right , i \}} R_{jk} \;\;  \cup
\;\; S  \]
\[ \begin{array}{lcl}
\Internal{Q} & = & Q \setminus \{ \qinit , \qfin \} \\
R_{rr} & = &  \{ (\qinit , u) \rightarrow (v, \qfin ) \mid (\qinit ,
r(u)) \rightarrow (r(v), \qfin ) \in R \} \\
  R_{rl} & = & \{ (\qinit , u) \rightarrow (v, \pinit ) \mid (\qinit ,
r(u)) \rightarrow (l(v), \qfin ) \in R \} \\
  R_{ll} & = &  \{ (\pfin , u) \rightarrow (v, \pinit ) \mid (\qinit ,
l(u)) \rightarrow (l(v), \qfin ) \in R \} \\
  R_{lr} & = & \{ (\pfin , u) \rightarrow (v, \qfin ) \mid (\qinit ,
l(u)) \rightarrow (r(v), \qfin ) \in R \} \\
  R_{ii} & = & \{ (q, u) \rightarrow (v, q') \in R \mid q, q' \in
\Internal{Q} \} \\
  R_{ri} & = & \{ (\qinit , u) \rightarrow (v, q) \mid (\qinit , r(u))
\rightarrow (v, q) \in R, q \in \Internal{Q} \}  \\
  R_{li} & = & \{ (\pfin , u) \rightarrow (v, q) \mid (\qinit , l(u))
\rightarrow (v, q) \in R, q \in \Internal{Q} \} \\
  R_{il} & = &  \{ (q , u) \rightarrow (v, \pinit ) \mid (q, u)
\rightarrow (l(v), \qfin ) \in R, q \in \Internal{Q} \} \\
  R_{ir} & = & \{ (q , u) \rightarrow (v, \qfin ) \mid (q, u)
\rightarrow (r(v), \qfin ) \in R, q \in \Internal{Q} \} 
\end{array} \]
\caption{Linear Application}
\end{figure*}

\noindent The key result we need is the following.
\begin{proposition}
(i) $\llbang f_{\Aut} = f_{\llbang \Aut}$. \\
(ii) $\LApp{f_{\Aut}}{f_{\BAut}} = f_{\LApp{\Aut}{\BAut}}$.
\end{proposition}

\noindent \proof \, \noindent (i)  $\llbang f_{\Aut}(p(t,u)) = p(t,v)$ iff
$f_{\Aut}(u) = v$ iff $u \labarrow{\Aut}^{\ast} v$ iff $p(t,u)
\labarrow{\llbang \Aut}^{\ast} p(t,v)$.

\noindent (ii) Let $\mathcal{C} = \LApp{\Aut}{\BAut}$.
Suppose $\LApp{f_{\Aut}}{f_{\BAut}}(t) = u$.
Then either $f_{rr}(t) = u$, or $f_{rl}(t)=v$, $g(v) = w_1$,
$f_{ll}(w_{1}) = w_2$, $g(w_{2})=w_3$, \ldots , $f_{ll}(w_{k}) =
w_{k+1}$, $g(w_{k+1}) = w_{k+2}$, $f_{lr}(w_{k+2}) = u$.
In the first case, $(\qinit , r(t)) \labarrow{\Aut}^{\ast} (\qfin , r(u))$,
and hence $(\qinit , t) \labarrow{\mathcal{C}}^{\ast} (\qfin , u)$.
In the latter case, $(\qinit , r(t)) \labarrow{\Aut}^{\ast} (\qfin ,
l(v))$,
$(\pinit , v) \labarrow{\BAut}^{\ast} (\pfin , w_{1})$,
$(\qinit , l(w_{1})) \labarrow{\Aut}^{\ast} (\qfin ,
l(w_{2}))$,
$(\pinit , w_{2}) \labarrow{\BAut}^{\ast} (\pfin , w_{3})$, \ldots ,
$(\qinit , l(w_{k})) \labarrow{\Aut}^{\ast} (\qfin ,
l(w_{k+1}))$,
$(\pinit , w_{k+1}) \labarrow{\BAut}^{\ast} (\pfin , w_{k+2})$,
$(\qinit , l(w_{k+2})) \labarrow{\Aut}^{\ast} (\qfin ,
r(u))$, and hence again $(\qinit , t) \labarrow{\mathcal{C}}^{\ast} (\qfin , u)$.
Thus $\LApp{f_{\Aut}}{f_{\BAut}} \subseteq f_{\LApp{\Aut}{\BAut}}$. The
converse inclusion is proved similarly. $\;\;\; \qed$

\subsection{Finitely describable partial involutions are Automatic}
Now suppose we are given a finite description $S$ of a partial
involution $f$. 
We  define a corresponding automaton $\Aut$:
\[ \Aut = (\{ \qinit , \qfin \} , \qinit , \qfin , R) \]
where 
\[ R \;\; = \; \bigcup_{(t, u) \in S} \{ (\qinit , t) \rightarrow (u,
\qfin ), \;\; (\qinit , u) \rightarrow (t, \qfin ) \} .
 \]
It is immediate that $f_{\Aut} = f$.

Note that $\Aut$ has no internal states, and all
its rules are of the above special form. These features are typical of the automata
corresponding to \emph{normal forms} in our interpretation of
functional computation.

\subsection{The Automatic Universe}
The results of the previous two sections yield the following Theorem as an
immediate consequence.
\begin{theorem}
\label{revclth}
$\mathcal{R}$ is an affine combinatory sub-algebra of $\Inj$, where the
carrier of $\mathcal{R}$ is the set of all $f_{\Aut}$ for biorthogonal 
automata $\Aut$. Moreover, $\mathcal{S} = \PInv \cap \mathcal{R}$ is
an affine
combinatory sub-algebra of $\mathcal{R}$.
\end{theorem}

\noindent Thus we obtain a subalgebra $\mathcal{S}$ of
$\mathcal{R}$, of partial involutions realized by biorthgonal
automata; and even these very simple behaviours are computationally
universal. Partial involutions can be seen as ``copy-cat strategies''
\cite{gfc}.

\subsection{Minimal requirements on $\Sigma$}
We now pause  briefly  to consider our choice of the particular
signature $\Sigma$. We could in fact eliminate the unary operators $l$ 
and $r$ in favour of two constants, say $a$ and $b$, and use the
representation
\[ \begin{array}{lcl}
l(t) & \equiv & p(a,t) \\
r(t) & \equiv & p(b,t) \\
p(t,u) & \equiv & p(\const , p(t,u)) . 
\end{array} \]
We can in turn eliminate $a$ and $b$, e.g. by the definitions
\[ a \equiv p(\const , \const ) \qquad b \equiv p(p(\const , \const ),
\const ). \]
So one binary operation and one constant---i.e. the pure theory of
binary trees---would suffice.

On the other hand, if our signature only contains unary operators and
constants, then pattern-matching automata can be simulated by ordinary 
automata with one stack, and hence are not computationally universal
\cite{Min}.

This restricted situation is still of interest. It suffices to
interpret $\mathbf{BCK}$-algebras, and hence the \emph{affine} 
$\lambda$-calculus \cite{Hin}. Recall that the $\mathbf{B}$ and
$\mathbf{C}$ combinators have the defining equations
\[ \begin{array}{lcl}
\BBB \cdot x \cdot y \cdot z = x \cdot (y \cdot z) \\
\CCC  \cdot x \cdot y \cdot z = x \cdot z \cdot y 
\end{array} \]
and that $\mathbf{BCK}$-algebras admit bracket abstraction for the affine $\lambda$-calculus, which is subject to the constraint that
applications $M \cdot N$ can only be formed if no variable occurs free 
in both $M$ and $N$. The affine $\lambda$-calculus is strongly
normalizing in a number of steps linear in the size of the initial
term, since $\beta$-reduction strictly decreases the size of the term.

We build a  $\mathbf{BCK}$-algebra over automata by using Linear
instead of standard application, and defining automata for the
combinators
$\BBB$, $\CCC$ and $\KK$ without using the binary operation symbol
$p$.
For reference, we give the set of transition rules for each of these
automata: \\
$R_{\KK}$ (linear version):
\[ r(r(x)) \leftrightarrow l(x) \]
$R_{\BBB}$:
\[ \begin{array}{lcl}
l(r(x)) & \leftrightarrow & r(r(r(x))) \\
l(l(x)) & \leftrightarrow & r(l(r(x))) \\
r(l(l(x))) & \leftrightarrow & r(r(l(x)))
\end{array} \]
$R_{\CCC}$:
\[ \begin{array}{lcl}
l(l(x)) & \leftrightarrow & r(r(l(x))) \\
l(r(l(x))) & \leftrightarrow & r(l(x))) \\
l(r(r(x))) & \leftrightarrow & r(r(r(x)))
\end{array} \]
Note that, since only unary operators appear in the signature, these
automata can be seen as performing \emph{prefix string rewriting}
\cite{KM89}.

\section{Compiling functional programs into reversible computations}
Recall that the pure $\lambda$-calculus is rich enough to represent
data-types such as integers, booleans, pairs, lists, trees, and
general inductive types \cite{GLT}; and control structures including recursion,
higher-order functions, and continuations \cite{Plo75}. A
representation of database query languages in the pure
$\lambda$-calculus is developed in \cite{HKM}. The
$\lambda$-calculus can be compiled into combinators, and in fact this
has been extensively studied as an implementation technique
\cite{PJ87}. Although combinatory weak reduction does not capture all of
$\beta$-reduction, it suffices to capture computation over
``concrete'' data types such as integers, lists etc., as shown e.g. by 
Theorem~\ref{clcomp}. Also, combinator algebras form the basic ingredient for
\emph{realizability constructions}, which are a powerful tool for
building models of very expressive type theories (for textbook
presentations see e.g. \cite{AL91,Cro93}). By our results in the previous section, a
combinator program $M$ can be compiled in a syntax-directed fashion into a 
biorthogonal automaton $\Aut$. Moreover, note that the size of $\Aut$ is \emph{linear}
in that of $M$.

It remains to specify how we can use $\Aut$ to ``read out'' the result 
of the computation of $M$. What should be borne in mind is that the
automaton $\Aut$ is giving a description of the \emph{behaviour} of
the functional process
corresponding to the program it has been compiled from. It is
\emph{not} the case that the terms in $\TSig$ input to and output from the
computations of $\Aut$ correspond directly to the inputs and outputs
of the functional computation. Rather, the input also has to be
compiled as part of the functional term to be evaluated---this is
standard in functional programming generally.\footnote{However, note that, by
compositionality, the program can be compiled once and for all into an
automaton, and then each input value can be compiled and
``linked in'' as required.} The automaton resulting
from compiling the program \emph{together with its input} can then be
used to deduce the value of the output, provided that the output is a
concrete value.

We will focus on \emph{boolean-valued} computations, in which the
result of the computation is either $\true$ or $\false$, which we
represent by the combinatory expressions $\KK$ and $\KK \cdot \II$
respectively. By virtue of the standard results on combinatory
computability such as Theorem~\ref{clcomp}, for any (total) recursive
predicate $P$, there is a closed combinator expression $M$ such that,
for all $n$, $P(n)$ holds if and only if
\[ \CL \vdash M \cdot \bar{n} = \KK , \]
and otherwise $\CL \vdash M \cdot \bar{n} = \KK \cdot \II$.
Let the automaton obtained from the term $M \cdot \bar{n}$ be
$\Aut$. Then by Theorem~\ref{revclth}, $f_{\Aut} = f_{\KK}$ or
$f_{\Aut} = f_{\KK \cdot \II}$. Thus to
test whether $P(n)$ holds, we run $\Aut$ on the input term $r(r(\const 
))$. If we obtain a result of the form $l(u)$, then $P(n)$ holds,
while if we obtain a result of the form $r(v)$, it does not.
Moreover, this generalizes immediately to predicates on tuples, lists, trees 
etc., as already explained.

More generally, for computations in which e.g. an integer is
returned, we can run a sequence of computations on the automaton
$\Aut$, to determine which value it
represents. Concretely, for Church numerals, the sequence would look
like this.
Firstly, we run the automaton on the input $r(r(\const ))$. If the
output has the form $r(l(u))$ (so that the term is `$\lambda f.\,
\lambda x.\, x$') then the result is $0$. Otherwise, it must have the
form $l(p(u,r(v)))$ (so it is of the form $\lambda f. \, \lambda x. \, 
f \ldots $, i.e. it is the successor of \ldots), and then we run the
automaton again on the input term $l(p(u,l(p(\const , v)))$. If we now 
get a response of the form $r(l(u ))$, then the result is the successor of
0, i.e. 1 (!!). Otherwise \ldots

In effect, we are performing a meta-computation (which \textit{prima
facie} is irreversible), each ``step'' of which is a reversible computation, to read out the output.
It could be argued that something analogous to this always happens in an implementation of 
a functional programming language, where at the last step the result 
of the computation has to be converted into human-readable output, and 
the side-effect of placing it on an output device has to be achieved.

This aspect of recovering the output deserves further attention, and
we hope to study it in more detail in the future.

\subsection*{Pure vs. Applied $\lambda$-calculus}
Our discussion has been based on using the pure $\lambda$-calculus or
CL, with no constants and $\delta$-rules \cite{HS,Bar84}. Thus
integers, booleans etc. are all to be represented as
$\lambda$-terms. The fact that $\lambda$-calculus and Combinatory
Logic can be used to represent data as well as control is an important 
facet of their universality; but in the usual practice of functional
programming, this facility is not used, and applied $\lambda$-calculi
are used instead. It is important to note that this option is \emph{not} open 
to us if we wish to retain reversibility. Thus if we extend the
$\lambda$-calculus with e.g. constants for the boolean values and
conditional, and the usual $\delta$-rules, then although we could
continue to interpret terms by orthogonal pattern-matching automata,
\emph{biorthogonality---i.e. reversibility---would be lost}. This can 
be stated more fundamentally in terms of Linear Logic: while the
multiplicative-exponential fragment of Linear Logic (within which the
$\lambda$-calculus lives) can be interpreted in a perfectly reversible 
fashion (possibly with the loss of soundness of some conversion rules \cite{Gi89,AHS}), this fails
for the additives. This is
reflected formally in the fact that in the passage from modelling the
pure $\lambda$-calculus, or Multiplicative-Exponential Linear Logic,
to modelling PCF, the property of partial injectivity of the functions 
$f_{\Aut}$ (the ``history-free strategies'' in \cite{gfc,pcf}) is
lost, and non-injective partial functions must be used
\cite{gfc,pcf,Mac}.
It appears that this gives a rather fundamental
delineation of the boundary between reversible and irreversible
computation in logical terms. This is also reflected in the
denotational semantics of the $\lambda$-calculus: for the pure
calculus, complete lattices  arise naturally as the canonical models
(formally, the property of being a lattice is preserved by
constructions such as function space, lifting, and inverse limit), while 
when constants are added, to be modelled by sums, inconsistency arises and the natural models
are cpo's \cite{lazy}. This suggests that the pure $\lambda$-calculus
itself provides the ultimate reversible simulation of the irreversible 
phenomena of computation.

\section{Universality}
A minor variation of the ideas of the previous section suffices to
establish universality of our computational model. Let $W$ be a
recursively enumerable set. There is a closed combinatory term $M$
such that, for all $n \in \Nat$,
\[ n \in W \;\; \Longleftrightarrow \;\; \CL \vdash M \cdot \bar{n} =
\bar{0} \]
and if $n \not\in W$ then $M \cdot \bar{n}$ does not have a normal
form. Let $\Aut$ be the automaton compiled from $M \cdot
\bar{n}$. Then we have a reduction of membership in $W$ to the
question of whether $\Aut$ produces an output in response to the input 
$r(r(\const ))$. As an immediate consequence, we have the following
result.
\begin{theorem}
Termination in biorthogonal automata is undecidable; in fact, it is
$\Sigma^{0}_{1}$-complete.
\end{theorem}

\noindent As a simple corollary, we derive the following result.
\begin{proposition}
Finitely describable partial involutions are not closed under linear
application.
\end{proposition}

\noindent \proof \ \   The linear combinators are all interpreted by finitely
describable partial involutions, and it is clear that replication
preserves finite describability. Hence if linear application also
preserved finite describability, all combinator terms would denote
finitely describable partial involutions. However, this would
contradict the previous Theorem, since termination for a finitely
describable partial involution reduces to a finite number of instances 
of pattern-matching, and hence is decidable. $\;\; \qed$

\noindent This leads to the following:
\begin{quotation}
\textbf{Open Question}: Characterize those partial involutions in
$\mathcal{S}$, or alternatively, those which arise as denotations of
combinator terms.
\end{quotation}
%\section{Concluding remarks}

\end{document}